\documentstyle[12pt]{article}

\expandafter\ifx\csname mathrm\endcsname\relax\def\mathrm#1{{\rm #1}}\fi

\makeatletter

\if@twoside \oddsidemargin -17pt \evensidemargin 00pt
\else \oddsidemargin 00pt \evensidemargin 00pt
\fi
\def\section{\@startsection {section}{1}{\z@}{+3.0ex plus +1ex minus
+.2ex}{2.3ex plus .2ex}{\normalsize\bf}}
\def\subsection{\@startsection{subsection}{2}{\z@}{+2.5ex plus +1ex
minus +.2ex}{1.5ex plus .2ex}{\normalsize\bf}}
\def\subsubsection{\@startsection{subsubsection}{3}{\z@}{+3.25ex plus
+1ex minus +.2ex}{1.5ex plus .2ex}{\normalsize\bf}}

\expandafter\ifx\csname mathrm\endcsname\relax\def\mathrm#1{{\rm #1}}\fi

\makeatother

\def\beq{\begin{equation}}
\def\eeq{\end{equation}}
\def\beqar{\begin{eqnarray}}
\def\eeqar{\end{eqnarray}}
\def\barr#1{\begin{array}{#1}}
\def\earr{\end{array}}
\def\bfi{\begin{figure}}
\def\efi{\end{figure}}
\def\btab{\begin{table}}
\def\etab{\end{table}}
\def\bce{\begin{center}}
\def\ece{\end{center}}
\def\nn{\nonumber}

\def\text{\textstyle}

\def\ga{\gamma}

\def\si{\sigma}

\def\refeq#1{\mbox{(\ref{#1})}}
\def\reffi#1{\mbox{Fig.~\ref{#1}}}

\def\citeres#1{\mbox{Refs.~\cite{#1}}}

\newcommand{\GeV}{\unskip\,\mathrm{GeV}}

\newcommand{\TeV}{\unskip\,\mathrm{TeV}}

\renewcommand{\O}{{\cal O}}
\newcommand{\Oa}{\mathswitch{{\cal{O}}(\alpha)}}

\def\mathswitchr#1{\relax\ifmmode{\mathrm{#1}}\else$\mathrm{#1}$\fi}

\newcommand{\PW}{\mathswitchr W}
\newcommand{\PZ}{\mathswitchr Z}

\newcommand{\Pe}{\mathswitchr e}
\newcommand{\Pne}{\mathswitch \nu_{\mathrm{e}}}

\newcommand{\Pep}{\mathswitchr {e^+}}
\newcommand{\Pem}{\mathswitchr {e^-}}

\newcommand{\PWm}{\mathswitchr {W^-}}

\def\mathswitch#1{\relax\ifmmode#1\else$#1$\fi}

\newcommand{\MW}{\mathswitch {M_\PW}}

\newcommand{\MZ}{\mathswitch {M_\PZ}}

\newcommand{\Me}{\mathswitch {m_\Pe}}

\newcommand{\sw}{\mathswitch {s_\PW}}

\def\solid{\raise.9mm\hbox{\protect\rule{1.1cm}{.2mm}}}
\def\dashed{$---$}
\def\dotted{${\cdot\cdot\cdot\cdot\cdot\cdot{}\!}$}

\def\draftdate{\relax}
\def\mda{\relax}
\def\mua{\relax}
\def\mla{\relax}
\def\mpar#1{\relax}
\def\draft{
\def\draftdate{\today}
\def\mpar##1{\marginpar{\hbadness10000\sloppy\boldmath\bf##1}%
                      \typeout{marginpar: \noexpand##1}\ignorespaces}
\def\mda{\mpar{\hfil$\downarrow$\hfil}}
\def\mua{\mpar{\hfil$\uparrow$\hfil}}
\def\mla{\marginpar[\boldmath\hfil$\rightarrow$\hfil]%
                   {\boldmath\hfil$\leftarrow $\hfil}%
                    \typeout{marginpar: $\leftrightarrow$}\ignorespaces}
}

\makeatletter
\let\@eqnsel = \hfil

\def\eqnarray{\stepcounter{equation}\let\@currentlabel=\theequation
\global\@eqnswtrue
\global\@eqcnt\z@\tabskip\@centering\let\\=\@eqncr
$$\halign to \displaywidth\bgroup\hskip\@centering
  $\displaystyle\tabskip\z@{##}$\@eqnsel&\global\@eqcnt\@ne
  \hskip 2\arraycolsep \hfil${##}$\hfil
  &\global\@eqcnt\tw@ \hskip 2\arraycolsep $\displaystyle\tabskip\z@{##}$\hfil
   \tabskip\@centering&\llap{##}\tabskip\z@\cr}
\makeatother

\newcommand{\egnwezeg}{$\Pem\ga\to\PWm\Pne$, $\Pem\PZ$, $\Pem\ga$}

\newcommand{\egezeg}{$\Pem\ga\to\Pem\PZ$, $\Pem\ga$}
\newcommand{\egnw}{\mathswitch{\Pem\ga\to\PWm\Pne}}
\newcommand{\egez}{\mathswitch{\Pem\ga\to\Pem\PZ}}
\newcommand{\egeg}{\mathswitch{\Pem\ga\to\Pem\ga}}
\newcommand{\egnwgezgegg}{$\Pem\ga\to\PWm\Pne\ga$,
                           $\Pem\PZ\ga$, $\Pem\ga\ga$}

\newcommand{\egnwg}{$\Pem\ga\to\PWm\Pne\ga$}
\newcommand{\egezg}{$\Pem\ga\to\Pem\PZ\ga$}

\def\born{{\mathrm{Born}}}
\def\weak{{\mathrm{weak}}}
\def\full{{\mathrm{full}}}

\def\CMS{{\mathrm{CMS}}}
\def\cut{\Delta\theta}

\begin{document}
\thispagestyle{empty}
\def\thefootnote{\fnsymbol{footnote}}
\setcounter{footnote}{1}
\null
\draftdate \hfill CERN-TH.6980/93
\vskip 1cm
\vfil
\begin{center}
{\Large \bf Gauge-Boson Production\\
in Electron-Photon Collisions%
\footnote{Contribution to the Proceedings of the Workshop on
{\it \Pep\Pem Collisions at 500\GeV: The Physics Potential},
Munich, Annecy, Hamburg, 20 November 1992 to 3 April 1993.}
\par} \vskip 2.5em
{\large
{\sc A.\ Denner} \\[1ex]
{\it CERN, Geneva, Switzerland} \\[2ex]
{\sc S.\ Dittmaier}  \\[1ex]
{\it Physikalisches Institut, University of W\"urzburg, Germany}
\par} \vskip 1em
\end{center} \par
\vskip 4cm
\vfil
{\bf Abstract:} \par
We summarize results for the processes \egnwezeg\ within
the electroweak Standard Model. We discuss the
essential features of the corresponding lowest-order cross-sections
and present numerical results for the unpolarized cross-sections
including the complete \Oa~virtual, soft-photonic, and hard-photonic
corrections. While at low energies the weak corrections are dominated
by the leading universal corrections,
at high energies we find large, non-universal corrections,
which arise from vertex and box diagrams involving non-Abelian gauge
couplings.
\par
\vskip 2cm
\noindent CERN-TH.6980/93 \par
\vskip .15mm
\noindent August 1993 \par
\null
\setcounter{page}{0}
\clearpage

\def\thefootnote{\fnsymbol{footnote}}
\setcounter{footnote}{1}
\null
\vskip 2.0em
\begin{center}
{\large \bf Gauge-Boson Production 
in Electron-Photon Collisions
\par} \vskip 2.0em  minus 0.5em
{\normalsize
{\sc A.\ Denner$^1$ and S.~Dittmaier$^2$
} \\[3.5ex]
\parbox{10.8cm}{\normalsize
{\it$^1$ Theory Division, CERN, Geneva, Switzerland} \\[1ex]
{\it$^2$ Physikalisches Institut, University of W\"urzburg, Germany} }
\par}
\end{center} \par
\vskip 3.0em

\renewcommand{\thefootnote}{\arabic{footnote}}
\setcounter{footnote}{0}
\section{Introduction}
\label{intro}

Apart from the precise investigation of \Pep\Pem~reactions, the next
linear collider will most likely also allow the study of $\Pem\ga$ and
$\ga\ga$ reactions. Thereby the high-energy photons are not only
naturally generated by beamstrahlung and bremsstrahlung, but can
be produced much more efficiently via the Compton backscattering
of laser photons off high-energy electrons. The relatively clean
environment of $\Pem\ga$ and $\ga\ga$ collisions will
enable further precision tests of the electroweak standard model,
complementary to the ones in $\Pep\Pem$
collisions. In view of electroweak physics the most important
processes are those of gauge-boson production, i.e.\
in $\Pem\ga$ collisions \egnwezeg. In particular
\PW-boson production allows an investigation of the non-Abelian
$\ga\PW\PW$ coupling widely independent of other couplings \cite{Ch91}.
\PZ-boson production and Compton scattering provide information
about possible anomalous couplings between the photon and
the \PZ\ boson \cite{Re82}. Moreover backward Compton scattering turns out
to be well suited as a luminosity monitor for $\Pem\ga$
collisions, in a way similar to forward Bhabha
scattering in \Pep\Pem~annihilation.

Accurate theoretical predictions for the processes \egnwezeg\
require the inclusion of radiative corrections (RCs) at least
in $\O(\alpha)$ since the statistical error of the corresponding
cross-sections in the picobarn range is expected to be at the per cent
level for a 500\GeV\ collider. Here we summarize the essential results
for these processes, which have been extensively discussed in
\citeres{egfb,egfbg}. In particular
we review the lowest-order cross-sections, the
weak RCs, and the full $\O(\alpha)$ corrections, thereby concentrating
on the unpolarized case.
Numerical results are presented for energies ranging from 10\GeV\
\looseness -1
to~2\TeV.

\section{Lowest-order cross-sections}

Using the conventions and results of
\citeres{egfb}, the unpolarized, differential
lowest-order cross-sections for vanishing electron mass
can be easily calculated to
\beqar
\left(\frac{d\sigma}{d\Omega}\right)_{\born}^{\egnw} &=&
\frac{\alpha^2(-u)(s-\MW^2)}{8\sw^2s^3(t-\MW^2)^2}
\left[(s-\MW^2)^2+(u-\MW^2)^2\right],
\nn\\
\left(\frac{d\sigma}{d\Omega}\right)_{\mathrm{Born}}^{\egez} &=&
\left[(g_{\Pe\Pe\PZ}^+)^2+(g_{\Pe\Pe\PZ}^-)^2\right]
\frac{\alpha^2(s-\MZ^2)}{4s^3(-u)}
\left[(s-\MZ^2)^2+(u-\MZ^2)^2\right],
\nn\\
\left(\frac{d\sigma}{d\Omega}\right)_{\mathrm{Born}}^{\egeg} &=&
\frac{\alpha^2}{2s^2(-u)}(s^2+u^2).
\eeqar
Here $s=E_\CMS^2$, $t=-(s-M_B^2)\sin^2(\theta/2)$, and
$u=-(s-M_B^2)\cos^2(\theta/2)$ with the centre-of-mass energy $E_\CMS$,
the mass of the outgoing boson $M_B$ and the scattering angle
between the incoming and outgoing fermion $\theta$.

\begin{figure}[t]
\begin{center}
\begin{picture}(15,9.6)
\put(1,-1)  {\includegraphics{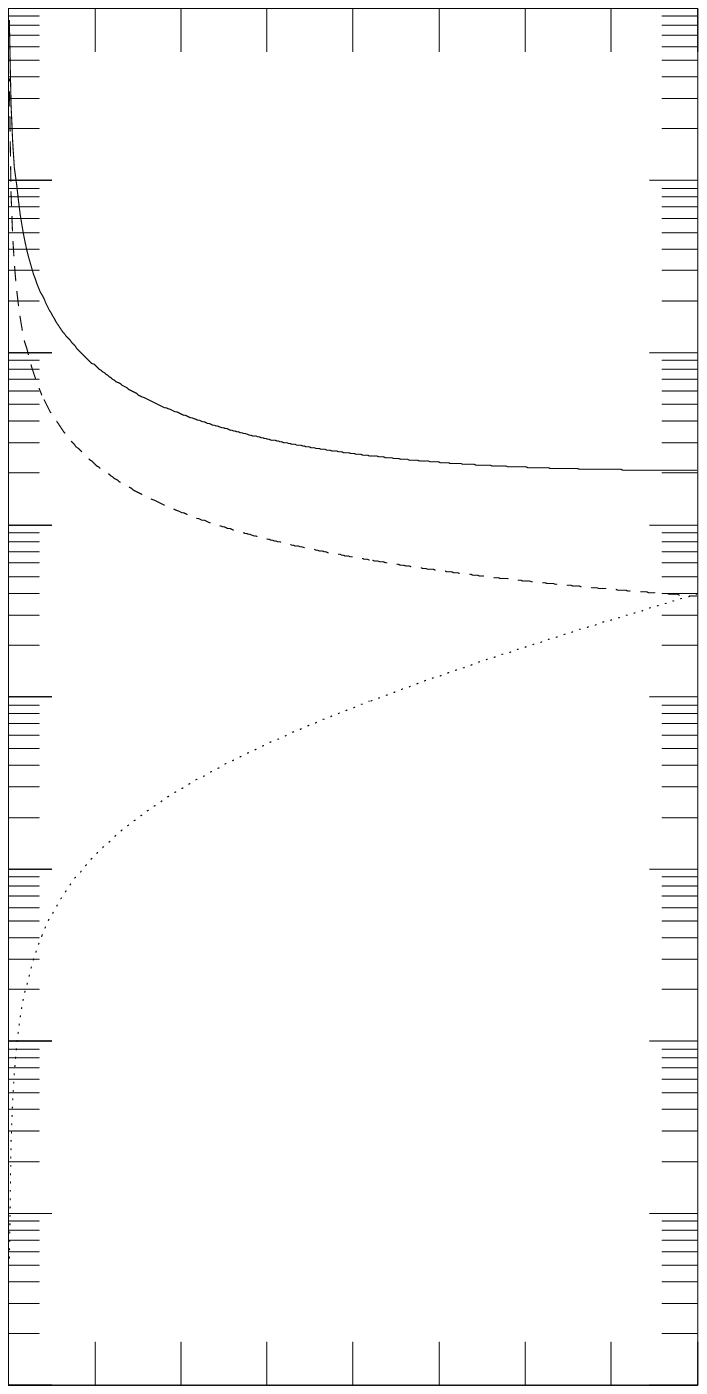}}
\put(5.5,-1){\includegraphics{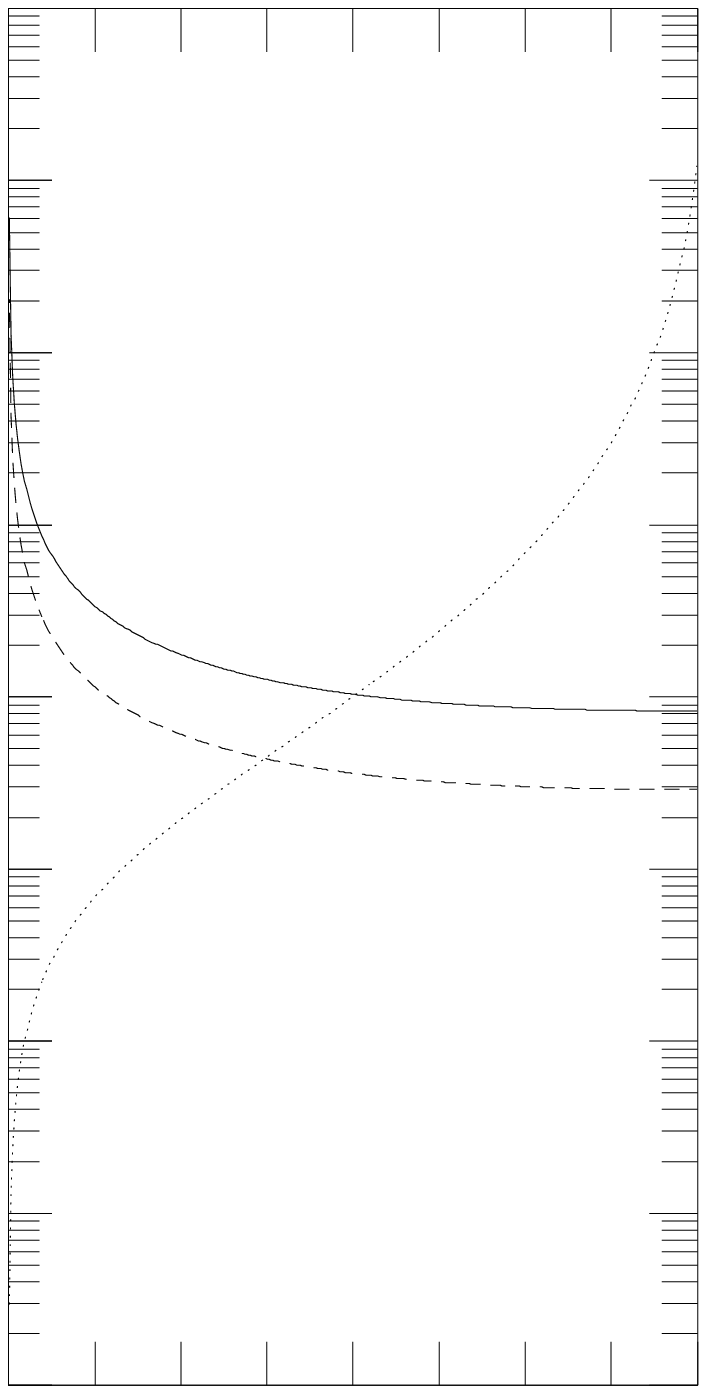}}
\put(10,-1) {\includegraphics{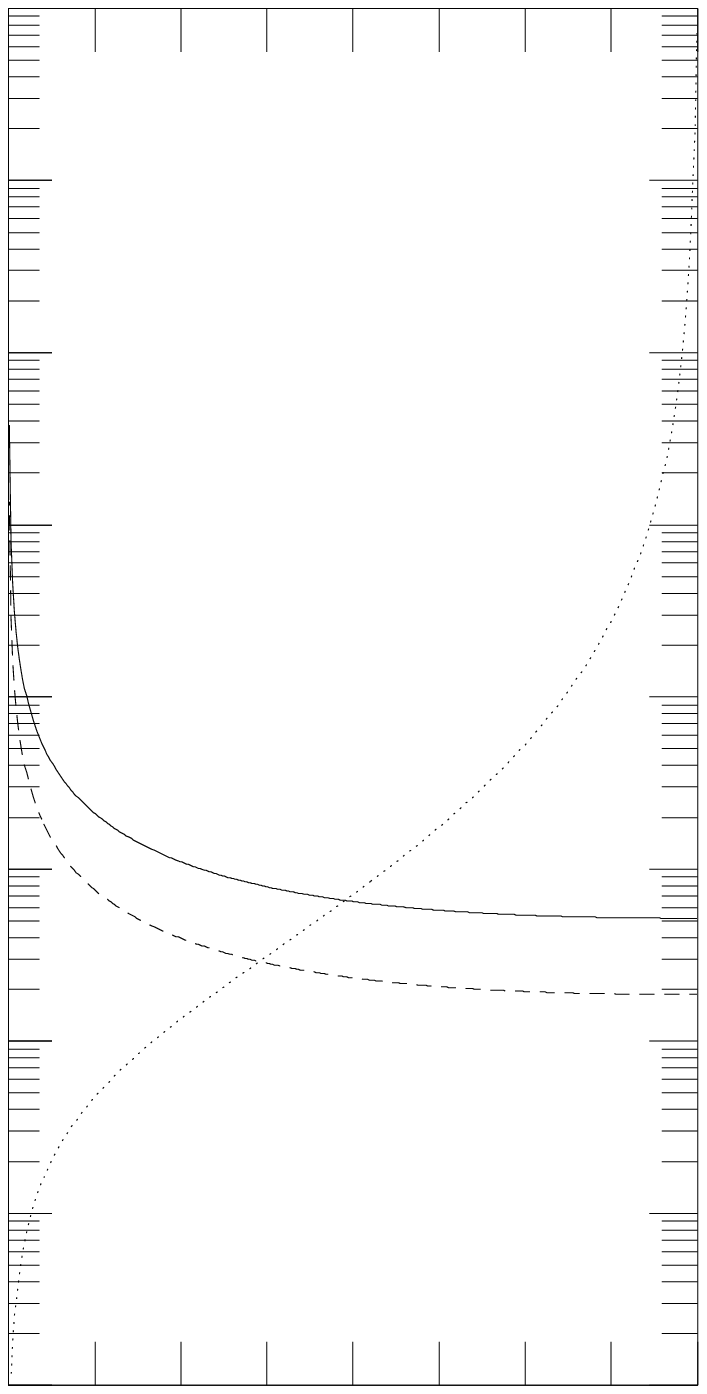}}
\put(-.5,5){$\displaystyle\frac{d\si}{d\Omega}/{\mathrm{pb}}$}
\put(.9,8.9){$10^{+3}$}
\put(.9,6.8){$10^{+1}$}
\put(.9,4.8){$10^{-1}$}
\put(.9,2.7){$10^{-3}$}
\put(.9,0.6){$10^{-5}$}
\put(2.5,9.3){$E_{\CMS}=100\GeV$}
\put(7,9.3){$E_{\CMS}=500\GeV$}
\put(11.7,9.3){$E_{\CMS}=2\TeV$}
\put(1.7,0.2){$-1\hspace{1.6cm}0\hspace{1.3cm}+1$}
\put(6.2,0.2){$-1\hspace{1.6cm}0\hspace{1.3cm}+1$}
\put(10.7,0.2){$-1\hspace{1.6cm}0\hspace{1.3cm}+1$}
\put(8,-.3){$\cos\theta$}
\end{picture}
\end{center}
\caption[]{Differential lowest-order cross-sections for unpolarized
particles:\\ {\solid~\protect\egeg}, \dashed~\protect\egez,
\dotted~\protect\egnw.}
\label{allbdiff}
\end{figure}

The angular dependence of the differential cross-sections,
illustrated in \reffi{allbdiff},
reflects the pole structure of single Feynman diagrams
contributing to the individual processes. In particular \egnw\
is characterized by a $t$-channel pole in the forward direction
and \egezeg\ are dominated by $u$-channel poles in the backward
direction. While the cross-section for \PW~production vanishes in the
backward direction, the ones for the production of neutral bosons
are finite everywhere.

\begin{figure}[t]
\begin{center}
\begin{picture}(15,8.6)
\put(-.2,-.5){\includegraphics{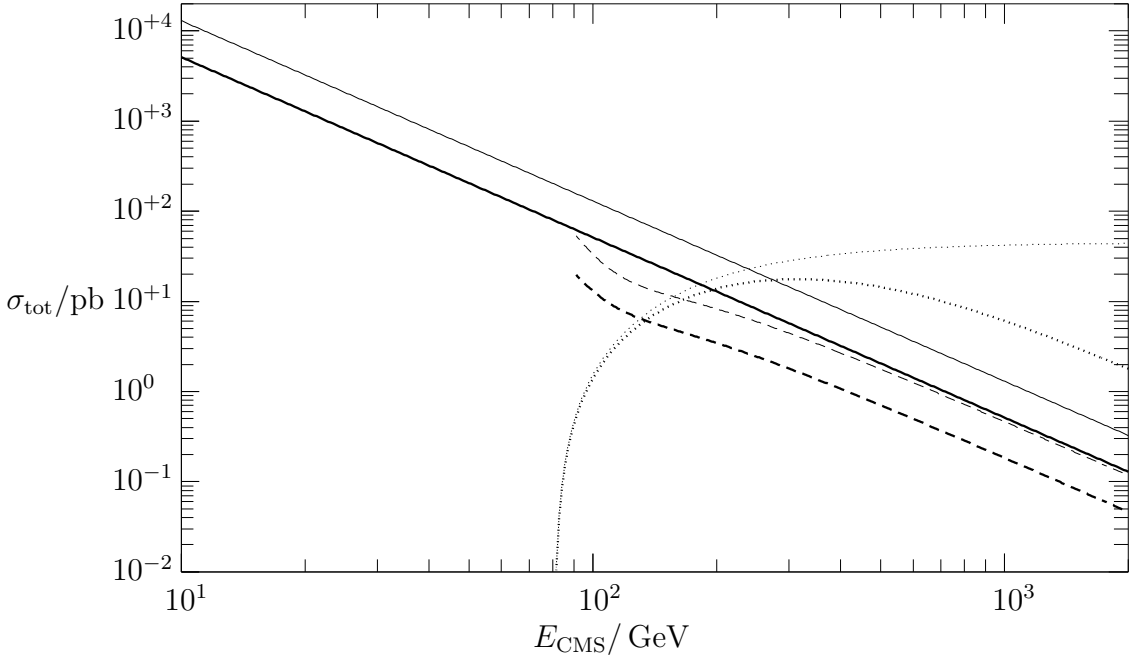}}
\put(0,4.5){$\si_{\mathrm{tot}}/{\mathrm{pb}}$}
\put(7,0){$E_{\CMS}/\GeV$}
\put(2.1,.5){$10^{1}$}
\put(7.6,.5){$10^{2}$}
\put(13.1,.5){$10^{3}$}
\put(1.4,8.1){$10^{+4}$}
\put(1.4,6.9){$10^{+3}$}
\put(1.4,5.7){$10^{+2}$}
\put(1.4,4.5){$10^{+1}$}
\put(1.4,3.3){$10^{0}$}
\put(1.4,2.1){$10^{-1}$}
\put(1.4,0.9){$10^{-2}$}
\end{picture}
\end{center}
\caption[]{Integrated lowest-order cross-sections for unpolarized
particles and two different cuts $\cut=20^\circ$ and $\cut=1^\circ$
($0^\circ$ for \egnw)
(same signature as in \protect\reffi{allbdiff}).}
\label{allb}
\end{figure}

The integrated lowest-order cross-sections are
\looseness -1
shown as functions of $E_{\CMS}$ in \reffi{allb}.
Owing to the (double) $t$-channel pole of
the differential cross-section, the total cross-section of \egnw\
approaches a constant in the high-energy limit. However, this
constant is replaced by a $1/s$ dependence
if the angular region around the beam axis is excluded from the phase
space. Such an angular cut $\Delta\theta <\theta <180^\circ-\Delta\theta$
does not only correspond to the experimental situation, it is
also necessary to avoid the backward singularity
($u$-channel pole) for \egezeg, which would be regularized by a
finite electron mass. The integrated cross-section
for \egez\ starts with a finite value at threshold since the phase-space
factor $(s-\MZ^2)$ in the numerator of differential cross-section
is cancelled exactly by the factor $(-u)$ in the denominator.
In contrast, the cross-section for \PW~production is suppressed by
$\beta^2$ ($\beta=(s-\MW^2)/(s+\MW^2) =$ velocity of \PW~boson)
close to threshold.
On the other hand W-boson production is dominating for energies above
200--300\GeV.
Compton scattering is preferred to \PZ-boson
production by roughly a factor 2--3, for all energies.

In \citeres{egfb} we have also discussed the various polarized
cross-sections. For all these processes the main contributions
are due to the channels where the helicities of the incoming
photon and the outgoing boson coincide. Relative to these,
the cross-sections for longitudinal \PW~and \PZ~production are
suppressed by $M_{\PW,\PZ}^2/s$. In the case of Compton
scattering the helicities of both the electron and the photon
are conserved in lowest order.

\section{Weak radiative corrections}

In the case of the neutral-current processes
\egezeg, the virtual pure QED corrections are formed by those
diagrams that involve virtual photons, the remaining gauge-invariant
subset of one-loop diagrams furnishes the weak
corrections. For the charged-current process
\egnw, the pure QED contributions
cannot be simply related to a subset of Feynman graphs. Instead
we define the relative weak RCs $\delta_{\mathrm{weak}}$ by
splitting off the leading QED corrections
\beq \label{QED}
\delta_{\mathrm{QED}}^{\Pem\ga\to\PWm\Pne}
=-\frac{\alpha}{\pi}\left\{
2\log\left(\frac{2\Delta E}{E_{\CMS}}\right)
\left[1+\log\left(\frac{\Me\MW}{\MW^{2}-u}\right)\right]
+\frac{3}{4}\log\left(\frac{\Me^{2}}{s}\right)\right\}
\label{nwlqed}
\eeq
from the complete virtual electroweak and real soft-photonic
correction $\delta$,
i.e.\
\beq
\delta_{\mathrm{weak}} = \delta - \delta_{\mathrm{QED}}.
\eeq
In \refeq{QED} $\Delta E$ denotes the soft-photon cut-off energy.

\begin{figure}
\begin{center}
\begin{picture}(15,8.8)
\put(.5,-1)  {\includegraphics{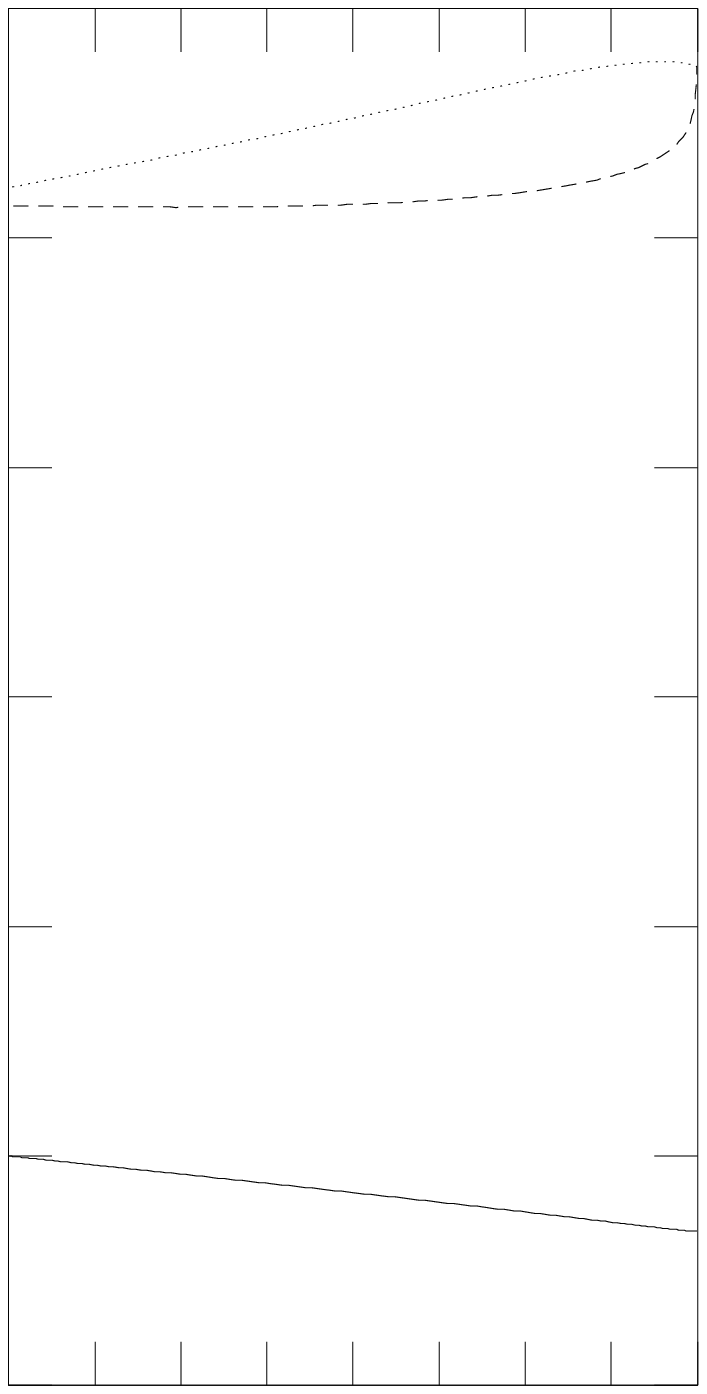}}
\put(5.5,-1) {\includegraphics{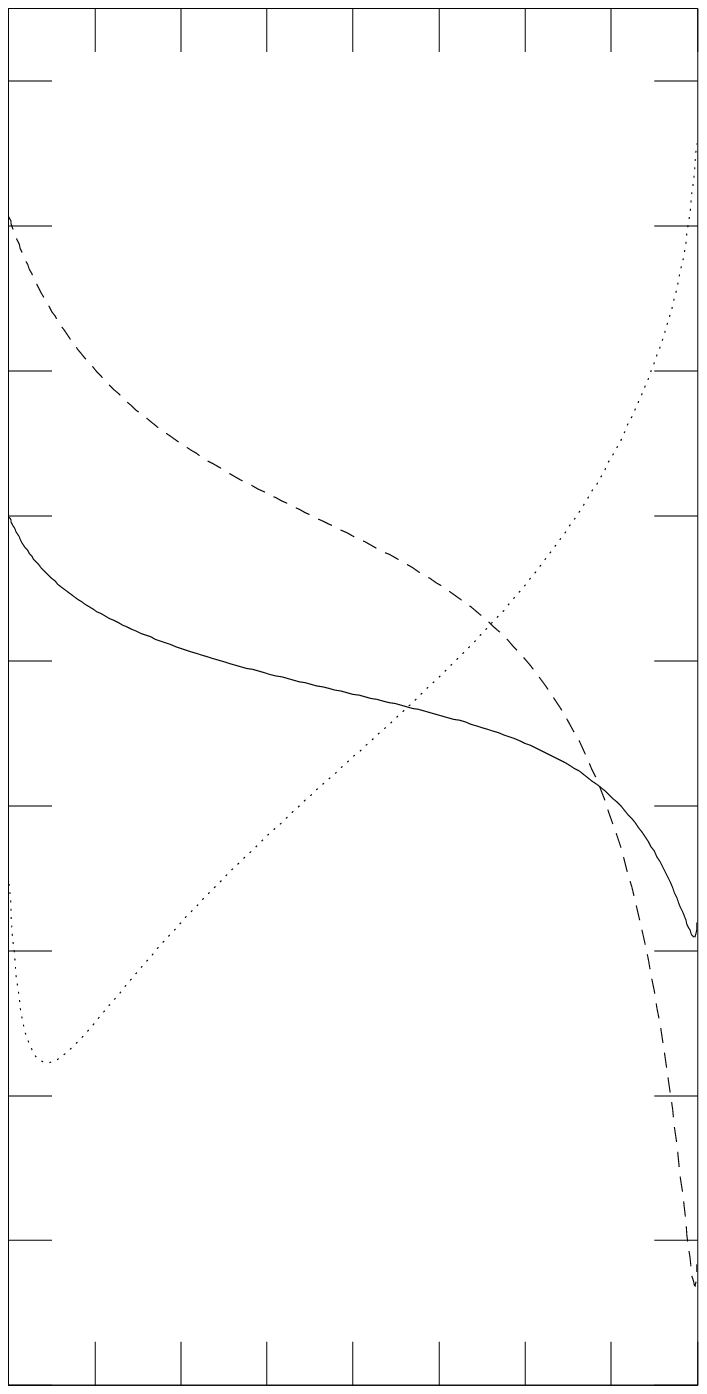}}
\put(10.5,-1){\includegraphics{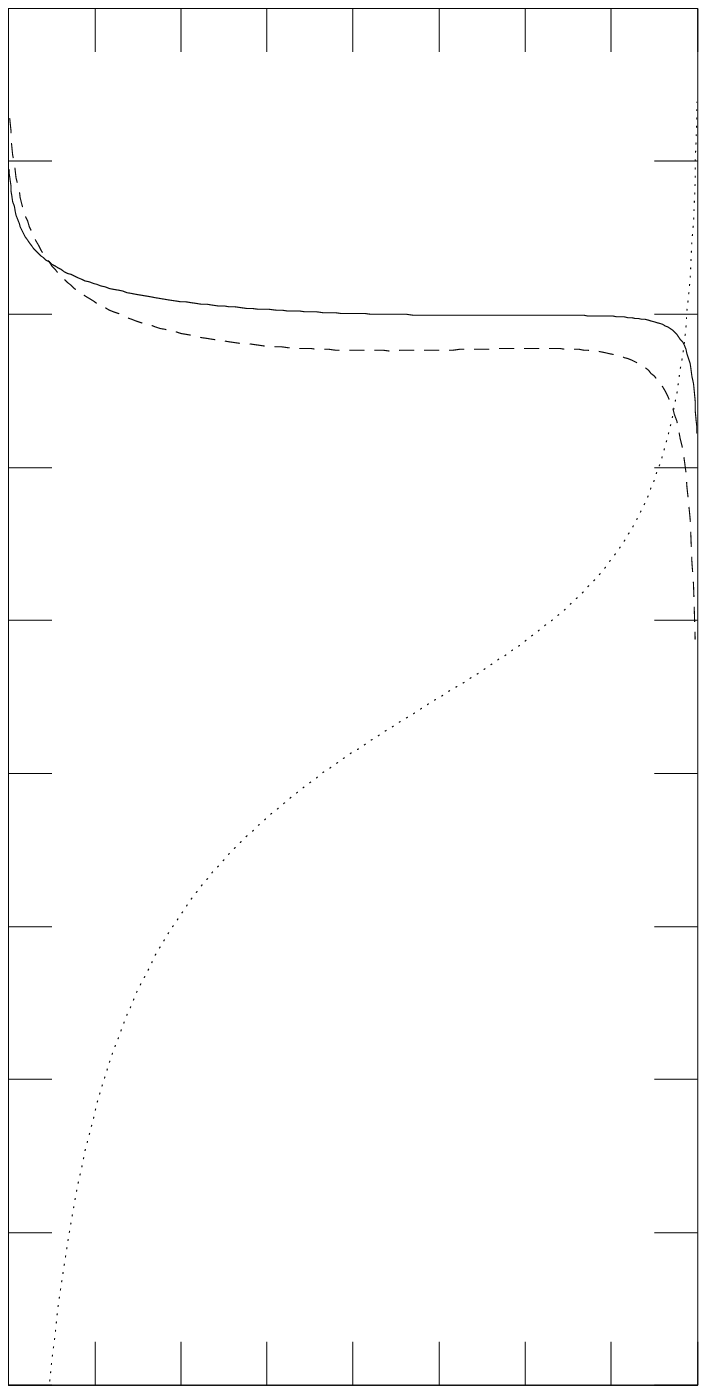}}
\put(-.6,5){$\delta_{\mathrm{weak}} /\%$}
\put(.9,8.1){$5$}
\put(.9,6.8){$4$}
\put(.9,5.6){$3$}
\put(.9,4.3){$2$}
\put(.9,2.9){$1$}
\put(.9,1.7){$0$}
\put(.6,0.5){$-1$}
\put(5.4,6.9){$\phantom{-1}4$}
\put(5.4,5.3){$\phantom{-1}0$}
\put(5.6,3.7){$-4$}
\put(5.6,2.1){$-8$}
\put(5.4,0.5){$-12$}
\put(10.3,7.3){$\phantom{-5}0$}
\put(10.4,5.5){$-20$}
\put(10.4,3.8){$-40$}
\put(10.4,2.1){$-60$}
\put(10.4,0.5){$-80$}
\put(1.8,8.5){$E_{\CMS}=100\GeV$}
\put(6.8,8.5){$E_{\CMS}=500\GeV$}
\put(12,8.5) {$E_{\CMS}=2\TeV$}
\put(1.1,0.2){$-1\hspace{1.5cm}0\hspace{1.2cm}+1$}
\put(6.1,0.2){$-1\hspace{1.5cm}0\hspace{1.2cm}+1$}
\put(11.1,0.2){$-1\hspace{1.5cm}0\hspace{1.2cm}+1$}
\put(8,-.3){$\cos\theta$}
\end{picture}
\end{center}
\caption{Weak corrections to the differential cross-sections for
unpolarized particles (same signature as in \protect\reffi{allbdiff}).}
\label{allwdiff}
\end{figure}

\begin{figure}
\begin{center}
\begin{picture}(15,8.6)
\put(-.2,-.5){\includegraphics{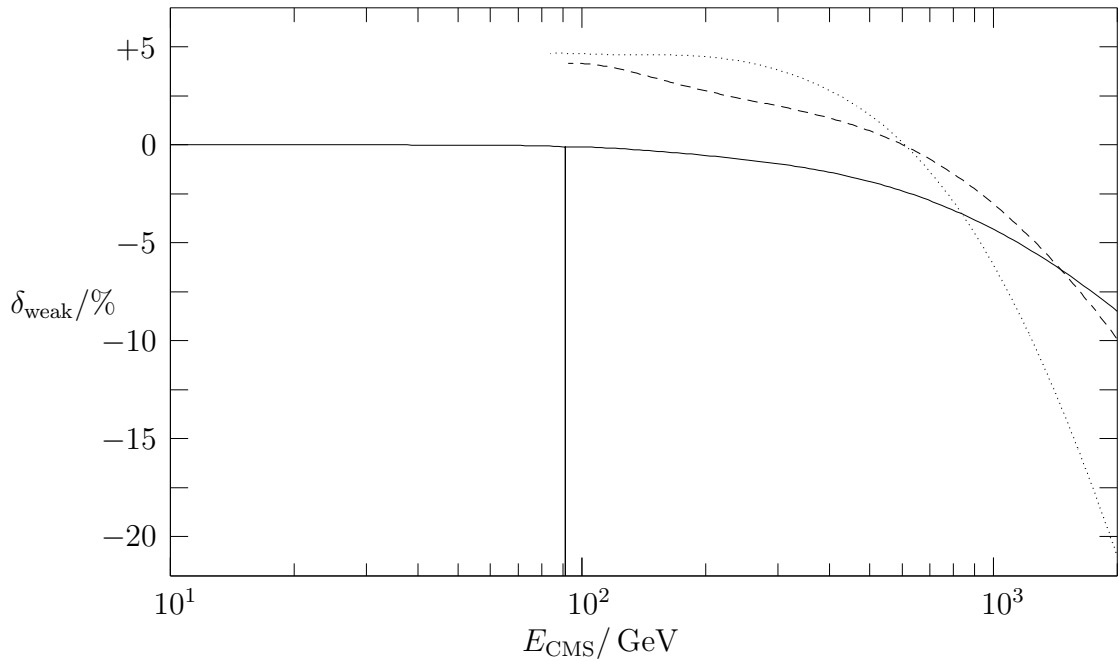}}
\put(0.2,4.5){$\delta_{\mathrm{weak}}/\%$}
\put(7,0){$E_{\CMS}/\GeV$}
\put(2.1,.5){$10^{1}$}
\put(7.6,.5){$10^{2}$}
\put(13.1,.5){$10^{3}$}
\put(1.6,7.9){$+5$}
\put(1.4,6.6){$\phantom{-1}0$}
\put(1.6,5.3){$-5$}
\put(1.4,4.0){$-10$}
\put(1.4,2.7){$-15$}
\put(1.4,1.4){$-20$}
\end{picture}
\end{center}
\caption[]{Weak corrections to the integrated cross-sections
($20^\circ <\theta <160^\circ$) for unpolarized particles
(same signature as in \protect\reffi{allbdiff}).}
\label{allw}
\end{figure}

{\samepage
Figure \ref{allwdiff} illustrates the angular dependence of
$\delta_{\mathrm{weak}}$ for the individual processes.
Note that the weak RCs vanish for Compton backscattering owing to the
}%
$u$-channel dominance and the usual definition of the electric charge
in the Thomson limit. For all the considered processes the angular
dependence of $\delta_{\mathrm{weak}}$ increases strongly with
energy. The large angular-dependent corrections at high energies
can be traced back to those vertex and box diagrams that contain
non-Abelian couplings. As a consequence the
{\it leading universal corrections}, i.e. the corrections associated
with the running of $\alpha$ and the $\rho$-parameter, which dominate
the weak corrections at low energies, are
definitely not sufficient at high energies. While $\delta_{\weak}$
gets very large in regions where the corresponding Born cross-sections
are suppressed, it is of the order of 10\% where the cross-sections
are sizeable. Therefore the weak RCs to the integrated cross-sections,
which are shown in \reffi{allw} as functions of $E_{\CMS}$,
are moderate at energies below 1\TeV.
In particular the integrated Compton cross-section
is practically not influenced by weak corrections
\looseness -1
for energies below the threshold singularity at $E_{\CMS}=\MZ$.

\section{Full $\O(\alpha)$ radiative corrections}

The full $\O(\alpha)$ corrections $\delta_\full$
include, in addition to the virtual and
real soft-photonic corrections, the hard-photonic ones
that correspond to the radiative processes \egnwgezgegg, with
bremsstrahlung photons of finite energy. The cross-sections of these hard
bremsstrahlung processes are discussed in \citeres{egfbg}.
Here we just mention that the main contributions arise from
collinear photon emission, where a finite electron mass has to be
taken into account to regularize the logarithmic collinear singularities.
This introduces finite-mass effects, which represent
the only origin for a possible spin flip of the fermions.

Figure \ref{fullrc} shows the full relative $\O(\alpha)$ corrections
to the single processes.
The bars indicate the statistical error of the Monte Carlo
integration over the three-particle phase space for the hard-%
bremsstrahlung cross-section, which is performed with VEGAS \cite{vegas}.
The phase space of the final state is restricted by the cuts%
\footnote{For Compton scattering this means that at least one emitted
photon has to fulfil these constraints.}%
\beq
\begin{array}[b]{rrcll}
& 20^\circ < & \theta'_\PW & < 160^\circ & \mbox{for \egnw}, \\
E'_\Pe > 0.2E,\quad &
20^\circ < & \theta'_\PZ, \theta'_\Pe & < 160^\circ &
\mbox{for \egez}, \\
E'_\ga, E'_\Pe > 0.2E,\quad &
20^\circ < & \theta'_\ga, \theta'_\Pe & < 160^\circ\quad &
\mbox{for \egeg},
\earr
\eeq
where all angles are relative to the incoming electron and primed
quantities refer to outgoing particles.
The $\O(\alpha)$ RCs to \egnwezeg\ turn out to be at the level
of several per cent for energies of several hundred GeV and increase for
higher energies $\propto \log^2s$.
In this limit the weak RCs exceed the photonic
corrections $\delta_{\mathrm{phot}}$ of pure QED, which grow only
logarithmically with energy.
For Compton scattering $\delta_{\mathrm{phot}}$ is of the form
\beq
\delta_{\mathrm{phot}}^{\Pem\ga\to\Pem\ga} =
C_1\log\left(\frac{\Me}{E_{\CMS}}\right)+C_2
\eeq
for fixed cut-off parameters, and $\delta_{\mathrm{weak}}$ is
responsible for the deviation of $\delta_{\mathrm{full}}$ from
a straight line in \reffi{fullrc}.
\begin{figure}
\begin{center}
\begin{picture}(15,8.6)
\put(-.2,-.5){\includegraphics{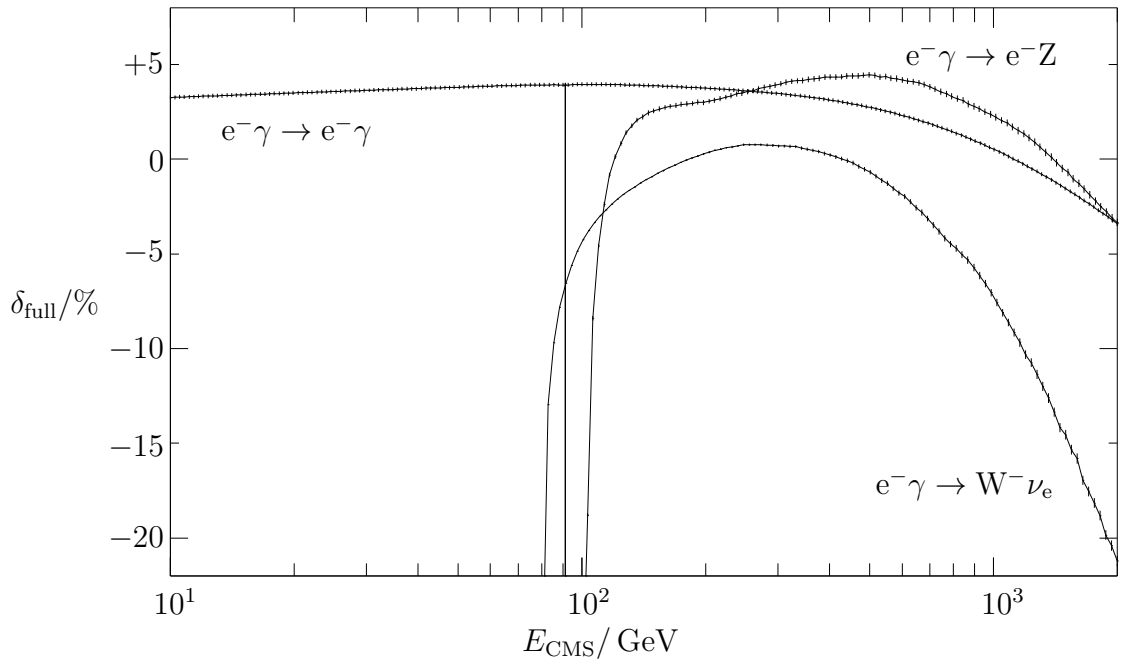}}
\put(0.2,4.5){$\delta_{\mathrm{full}}/\%$}
\put(7,0){$E_{\CMS}/\GeV$}
\put(3.0,6.8){\egeg}
\put(12.1,7.8){\egez}
\put(11.7,2.1){\egnw}
\put(2.1,.5){$10^{1}$}
\put(7.6,.5){$10^{2}$}
\put(13.1,.5){$10^{3}$}
\put(1.7,7.7){$+5$}
\put(1.5,6.4){$\phantom{-1}0$}
\put(1.7,5.2){$-5$}
\put(1.5,3.9){$-10$}
\put(1.5,2.6){$-15$}
\put(1.5,1.4){$-20$}
\end{picture}
\end{center}
\caption[]{Full \protect $\O(\alpha)$ corrections to the integrated
cross-sections for unpolarized particles.}
\label{fullrc}
\end{figure}

\section{Conclusion and outlook}

{\samepage
We have calculated the complete \Oa~Standard Model
radiative corrections, including
soft- and hard-photonic bremsstrahlung, to the processes \egnwezeg\
for arbitrary polarizations of the fermions and bosons.
Where the cross-sections are sizeable, the corrections are
typically at the level of a few per cent.
Close to threshold we find large corrections
for \egnw, $\Pem\PZ$ arising from soft photons. As a consequence these
corrections should be exponentiated. Moreover finite-width effects
should be included and a convolution with a realistic photon spectrum
should be performed. We do, however, not expect that these refinements
will drastically affect the size of the \Oa~corrections.

}


\begin{thebibliography}{9}
\frenchspacing
\newcommand{\zpC}[3]{{\sl Z. Phys.} {\bf C#1} (19#2) #3}
\newcommand{\npB}[3]{{\sl Nucl. Phys.} {\bf B#1} (19#2) #3}
\newcommand{\plB}[3]{{\sl Phys. Lett.} {\bf B#1} (19#2) #3}
\newcommand{\prD}[3]{{\sl Phys. Rev.} {\bf D#1} (19#2) #3}
\newcommand{\fp}[3]{{\sl Fortschr. Phys.} {\bf #1} (19#2) #3}
\newcommand{\cpc}[3]{{\sl Comp. Phys. Comm.} {\bf #1} (19#2) #3}
\newcommand{\nim}[3]{{\sl Nucl. Instr. Meth.} {\bf #1} (19#2) #3}
\newcommand{\jcp}[3]{{\sl J. Comp. Phys. } {\bf #1} (19#2) #3}


\bibitem{Ch91}
S.Y.\ Choi and F.\ Schrempp,
in: {\it Proceedings of the Workshop on
$\Pe^+\Pe^-$ Collisions at 500\,GeV: The Physics Potential},
Munich, Annecy, Hamburg, 1991, ed. P.~Zerwas,
DESY 92-123B (Hamburg, 1992), p. 793 and references therein;\\
G.~B\'elanger and F.~Boudjema, ibid., p.~783 and references therein;\\
O.\ Philipsen, \zpC{54}{92}{643}.

\bibitem{Re82}
F.M. Renard, \zpC{14}{82}{209}; \npB{196}{82}{93};\\
R. Ryzak, \npB{289}{89}{301}.

\bibitem{egfb}
A.~Denner and S.~Dittmaier, \npB{398}{93}{239} and 265;\\
CERN-TH.6876/93, to appear in {\sl Nucl. Phys.} {\bf B}.

\bibitem{egfbg} M.~B\"ohm and S.~Dittmaier,
{\it The hard bremsstrahlung process \egnwg}, W\"urzburg preprint
PRINT-93-0440 (1993), to appear in {\sl Nucl. Phys.} {\bf B};
{\it The hard bremsstrahlung process \egezg}, W\"urzburg preprint 1993;
\\ S.~Dittmaier, Dissertation, University of W\"urzburg, 1993.

\bibitem{vegas} G.P.\ Lepage, \jcp{27}{78}{192};
 Cornell Univ.~prep., CLNS-80/447 (1980).

\end{thebibliography}
\end{document}